\documentclass{article}

\usepackage[a4paper, margin=2cm]{geometry}

\usepackage{amsthm,amsfonts,amsmath,amssymb}
\usepackage{mathtools}
\usepackage{float}
\usepackage{bm}
\usepackage{verbatim}

\newcommand{\mb}[1]{\bm{#1}}
\newcommand{\pd}{\partial}

\DeclareMathOperator{\sign}{sign}

\usepackage{graphicx}
\graphicspath{{../fig/}{fig/}{..}}

\newcommand{\key}[1]%
{\,\raisebox{0mm}{\protect\includegraphics{key/#1}}\,}

\usepackage{xstring}
\newcommand{\figbf}[1]{\textbf{#1}}

\usepackage{xcolor}

\usepackage{algpseudocode}
\algnewcommand\True{\texttt{True}}
\algnewcommand\False{\texttt{False}}
\algnewcommand\Not{\textbf{not}}

\usepackage{hyperref}
\usepackage{nameref}
\usepackage{cleveref}
\crefname{figure}{Fig.}{Fig.}
\crefname{table}{Table}{Tables}
\crefname{equation}{}{}
\crefname{section}{Section}{Sections}
\crefname{appendix}{Appendix}{Appendix}

\usepackage{lineno}

\usepackage{subfiles}

\usepackage{xr}

\makeatletter
\newcommand*{\addFileDependency}[1]{% argument=file name and extension
  \typeout{(#1)}
  \@addtofilelist{#1}
  \IfFileExists{#1}{}{\typeout{No file #1.}}
}
\makeatother

\newcommand*{\myexternaldocument}[2]{%
    \externaldocument[#2]{#1}%
    \addFileDependency{#1.tex}%
    \addFileDependency{#1.aux}%
}

\IfFileExists{supp.aux}{
  \myexternaldocument{supp}{supp_}
}{
  \myexternaldocument{supp_pre}{supp_}
}

\newcommand{\suppreftable}[1]{Supplementary Table~\ref{supp_#1}}
\newcommand{\suppreffigure}[1]{Supplementary Figure~\ref{supp_#1}}
\newcommand{\suppnameref}[1]{\nameref{supp_#1}}

\title{Computing foaming flows across scales: \\ from  breaking waves to microfluidics}

\usepackage{authblk}

\author[a]{Petr Karnakov}
\author[a]{Sergey Litvinov}
\author[a,b,$\ast$]{Petros Koumoutsakos}
\affil[a]{Computational Science and Engineering Laboratory, ETH Zurich, Switzerland}
\affil[b]{School of Engineering and Applied Sciences, Harvard University, Cambridge, MA 02138, USA}
\affil[$\ast$]{corresponding author: petros@seas.harvard.edu}

\begin{document}

\maketitle

\begin{abstract}
%+% abs.tex
Crashing ocean waves, cappuccino froths and microfluidic bubble crystals are examples of foamy flows. Foamy flows are critical in numerous  natural and industrial processes and remain notoriously difficult to compute as they involve coupled, multiscale physical processes. Computations need to resolve the interactions of the bubbles with the fluid and complex boundaries, while  capturing the drainage and rupture of the microscopic liquid films at their interface.
We present a novel multilayer simulation framework (Multi-VOF)  that advances the state of the art in simulation capabilities of foamy flows. The framework  introduces a novel scheme for the
  distinct handling of multiple neighboring bubbles and a new regularization method that produces sharp interfaces and removes spurious fragments. Multi-VOF is verified and validated with experimental results and complemented with open source, efficient scalable software. 
We demonstrate capturing of bubble crystalline structures   in realistic  microfluidics devices and foamy flows involving tens of thousands of bubbles in a waterfall.
The present multilayer framework extends the classical volume-of-fluid
methodology and allows for unprecedented large scale, predictive simulations of flows with multiple interfaces.

%-% abs.tex
\end{abstract}

% keywords
%coalescence prevention, surface tension, volume-of-fluid, foam, embedded boundaries

%+% intro.tex
\section{Main}

Bubbly foams are the signature of violent crashes of ocean waves and happy
times accompanied by champagne. Bubble coalescence and collapse is central to
new products and technologies in areas ranging from food, to cosmetics and drug
delivery through precision microfluidics for emulsions and
foams~\cite{Stoffel2012,Anna2016,Prosperetti2017}.
Drainage and rupture of the liquid film separating bubbles leads to  their coalescence.
Surfactants~\cite{del2011} and other impurities, such as
electrolytes~\cite{craig1993}, in the liquid can delay or prevent coalescence.
Bubbles in clean liquids can also collide without coalescence if the film
drainage time is sufficiently large~\cite{chan2011}. 
Foams are composed by many bubbles separated by stable liquid films.
Foams are structural elements of insect, fish
and frog nests and central to numerous industrial processes and
medicine~\cite{Stone2009,Hill2017,Dollet2019}.

The simulation of foamy flows, involving  non-coalescing bubbles, presents a
number of formidable challenges in addition to those  associated with
resolving the bubble interactions with the fluid flow and the solid boundaries~\cite{Tryggvason2010}.
In foams the thickness of the film is usually several orders of magnitude
smaller than the size of bubbles.
In the limit of zero film thickness, the evolution of (dry) foams can be
predicted using Lagrangian techniques that track the surfaces between bubbles
with multiple junctions~\cite{brakke1992}. State of the art techniques involve
Eulerian methods such as the Voronoi Implicit Interface Method
(VIIM)~\cite{saye2011} that uses a single level-set function for all
interfaces.
More recently, methods such as the Lattice Boltzmann ~\cite{montessori2020}
have captured the effect of thin films using an empirical collision potential
but resort to mesoscale models with limited density ratio and artificial
compressibility.
In simulations of bubble collisions in turbulent flows~\cite{fang2017}, a short
range repulsive force between bubbles has been used to increase the surface
tension near the contact.
However, this can only remedy rapid collisions but not account for  stable
multi-bubble structures with thin films as in foams.

The classical volume-of-fluid (VOF) methodology \cite{Hirt1981,Prosperetti2009} has found
widespread success in simulations of engineering flows involving interfaces.
However, VOF leads to spurious coalescence of bubbles that approach each other
at a distance below one computational cell.
Such spurious coalescence can be prevented by introducing distinct volume
fraction fields for each bubble in the multi-marker volume-of-fluid~\cite{balcazar2015} or distinct
functions in level-set methods~\cite{coyajee2009}.
This prevention of coalescence has a computational cost that is proportional to
the number of bubbles in the simulation
(i.e.~$\mathcal{O}(N_\mathrm{bubbles}\,N_\mathrm{cells})$), and it is
prohibitive  for systems with a few hundred bubbles even in today's computer
architectures.
Furthermore as each distinct volume fraction field corresponds to  only one
bubble keeping track of all volume fraction fields is redundant. 
We propose a method to store multiple fields in a compact way
so that the number of the necessary scalar fields is constant,
independent on the number of bubbles in the simulation
(see~\nameref{s_methods}).
This technique is combined with VOF to simulate flows with non-coalescing bubbles.
The proposed  multilayer volume-of-fluid method (Multi-VOF) overcomes the above
mentioned challenges and allows for predictive simulations of flows involving
thousands of interacting and non-coalescing bubbles.  Multi-VOF requires only a
fixed number of fields,  and its computational cost is independent on the
number of bubbles in the simulation.

%-% intro.tex

\section{Results}
\label{s_results}

%+% meancurvflow.tex

\subsection{Constrained mean curvature flow, comparison with VIIM}
\label{s_meancurvflow}

\begin{figure}
  \centering
  \includegraphics{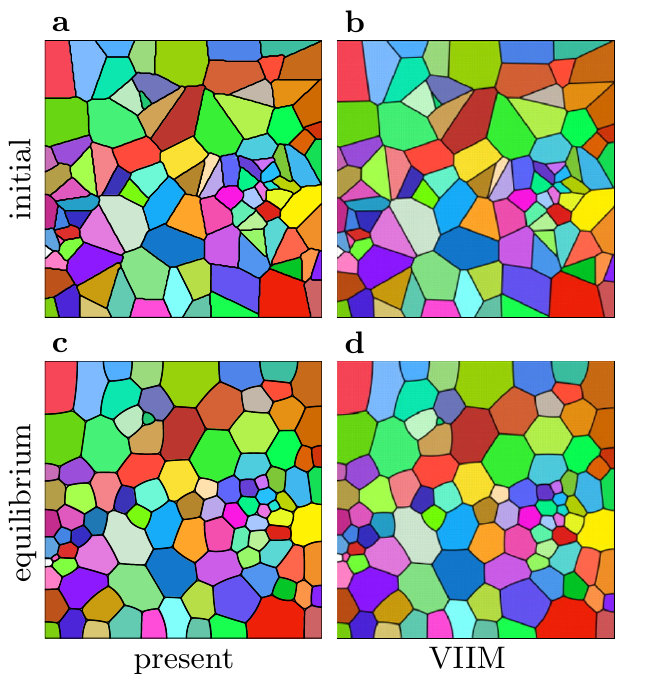}

  \caption{\fontsize{9}{9}\selectfont
    Constrained mean curvature flow.
    The initial field (\figbf{a,b}) is a Voronoi diagram.
    Multi-VOF method (\figbf{c}) produces results similar to that of
    the Voronoi implicit interface method (VIIM)~\cite{saye2011} (\textbf{d})
    on the same mesh of~$256^2$ cells.
    Interfaces at triple junctions form $120^\circ$ angles.
  }
  \label{f_meancurvflow}
\end{figure}

The capabilities of Multi-VOF are first assessed in  the limiting case of fluid flows with surface tension but no inertia. In this case the governing equations 
(see ~\nameref{s_methods})
simplify to a single  equation for the velocity field
$\mb{u} = 
\kappa\nabla \alpha - \nabla p$,
where the scalar field~$p(\mb{x},t)$ is used to ensure that the  velocity field is divergence-free ($\nabla\cdot \mb{u} = 0$). 
We use this velocity field with the advection equation
to compare Multi-VOF with the pioneering work in  \cite{saye2011}.
The initial conditions are adopted  from~\cite{saye2011}
and represent a Voronoi diagram of a randomly chosen set of 100 points with 
homogeneous Neumann boundary conditions for the volume fraction fields.
Initially, the interfaces  are straight lines and form multiple junctions at arbitrary angles.
As time evolves, only triple junctions remain and the angles between
the lines approach~$120^\circ$.
In~\cref{f_meancurvflow}, we compare the solution by our method
and the Voronoi Implicit Interface Method~\cite{saye2011}.
For Multi-VOF, dry foams are a limiting case
since there is no special treatment of triple junctions leading to the formation of small voids near the junctions.
Nevertheless, the results are of Multi-VOF and VIIM are in excellent agreement for the same mesh size. Applying penalization techniques can further improve the results
by removing the voids.

%-% meancurvflow.tex

%+% crystal.tex
\subsection{Microfluidic crystals}
\label{s_crystal}

Bubbles and droplets in microfluidic devices
organize into lattices called microfluidic crystals~\cite{beatus2006}.
They serve as prototypes of foam structures, compartments for chemical
reactions, and parts in production of
metamaterials~\cite{marmottant2009}.
A recent study~\cite{montessori2020} applied a mesoscale
lattice Boltzmann (LB) model to capture these flow structures.
The LB model includes a short range forcing term
that describes the combined effect of surface tension and near-contact
interactions to prevent coalescence.
The authors compare their results with experimental
data~\cite{marmottant2009} on foams of air bubbles in water.
However, due to the limited density ratio of the LB approach,
their mesoscale model is instead tuned for water drops embedded in oil.

Using the Multi-VOF method, we aim to reproduce the experimental study~\cite{raven2009}
on the formation of microfluidic crystals of bubbles in water.
Limited by numerical stability and computational cost,
we use lower values of the density ratio,
surface tension, viscosity and the channel length.
The device is based on a flow-focusing geometry~\cite{garstecki2005mechanism} and
consists in a planar network of rectangular ducts of height~$H$
with three inlets.
The gas is injected from one inlet at a fixed pressure~$P_g$
relative to the outlet pressure,
and the liquid comes through the other two with a total flow rate~$Q_l$.
The gas enters the channel through a contracting duct
that ends with an orifice of a width~$W_0$ expanding into the collection channel.
Walls of the channel are no-slip boundaries.
Parameters of the simulation are given in~\suppreftable{t_crystal}.

Bubbles in such devices are generated by the breakup of the
air thread in the inlet channel.  The period of breakup is determined
by the liquid flow rate and not by capillary time scales despite an apparent
similarity to the Rayleigh-Plateau instability~\cite{garstecki2005mechanism}.
In simulations, we forcibly separate the air thread at a regular interval~$T_b$.
Unless stated otherwise, the period of breakup equals $T_b=0.5\,W_0^2H/Q_l$
estimating the time it takes for the liquid to fill
the volume of a cavity forming right before the breakup.

\begin{figure}
  \centering
  \includegraphics{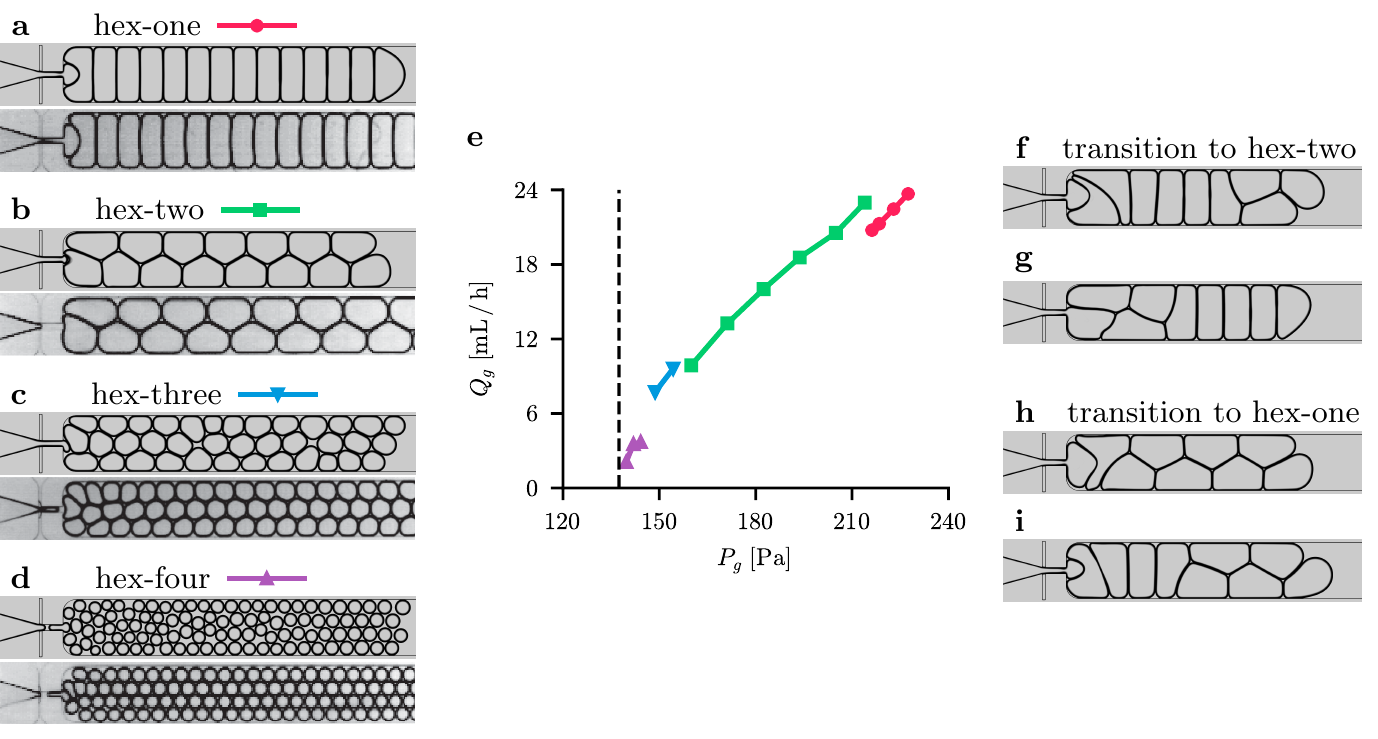}

  \caption{
    (\figbf{a-d})~%
    Crystalline structures of bubbles from simulations
    compared to experimental images~\cite{raven2009}.
    Hex-one ($P_g=216\;\mathrm{Pa}$, \figbf{a}),
    hex-two ($P_g=194\;\mathrm{Pa}$, \figbf{b}),
    hex-three ($P_g=149\;\mathrm{Pa}$, \figbf{c}),
    and hex-four ($P_g=140\;\mathrm{Pa}$, \figbf{d}).
    (\figbf{e})~%
    Gas flow rate as a function of the inlet gas pressure.
    The flow starts once the pressure exceeds the
    capillary threshold~\key{200}.
    Lines correspond to hex-one~\key{111},
    hex-two~\key{122}, hex-three~\key{133}, and hex-four~\key{144}
    structures.
    (\figbf{f-g})~%
    Spontaneous transitions between hex-one and hex-two
    with~$L=5.3\;\mathrm{mm}$ and $P_g=225\;\mathrm{Pa}$.
    Snapshots are taken at times
    $443\;\mathrm{ms}$ (\figbf{f}), $460\;\mathrm{ms}$ (\figbf{g}),
    $512\;\mathrm{ms}$ (\figbf{h}), and $529\;\mathrm{ms}$ (\figbf{i}).
  }
  \label{f_crystal}
\end{figure}

To obtain various crystalline structures, we vary the inlet gas pressure~$P_g$.
The gas flow rate~$Q_g$ and the bubble volume~$V_b$ as functions of pressure
are plotted in~\cref{f_crystal}.
The gas flow starts as soon as the pressure exceeds the capillary threshold
estimated as~$P_\mathrm{cap}=\sigma \big(\frac{1}{H} + \frac{1}{W_0}\big)$.
Raising the pressure enhances the gas flow rate~$Q_g$ and,
since the breakup period~$T_b$ is kept constant, increases the size of bubbles.
Each simulation with a given~$P_g$ is advanced
until equilibration, the evolution of the gas flow rate
for selected values of~$P_g$ is shown in~\suppreffigure{f_crystal_volume}.
Closely packed bubbles form~$120^\circ$ angles at triple junctions, and
junctions at the walls are~$90^\circ$.
Depending on the size, the bubbles organize into regular structures,
or flowing crystals. The structures are
named by the number of bubbles that fit in the channel
width~\cite{raven2009}: hex-one, hex-two, hex-three and so on.
Examples of the structures together with experimental images
are shown in~\cref{f_crystal} and Movies~S1-S4.
Stability of each structure is dictated by the corresponding value of the
surface energy. Smaller bubbles transition to higher-order structures.

The dissipation in foam is
proportional to its \textit{wetting perimeter},
i.e. the length of the menisci that confine the liquid
between the interface and the channel walls~\cite{cantat2004}.
Hex-one bubbles dissipate more than hex-two bubbles
and correspond to a slower flow~\cite{raven2009}.
Our simulations capture this effect as seen from~\cref{f_crystal} and Movie~S5.
If the structure remains the same,
increasing the pressure leads to a faster flow.
Conversely, if the structure changes,
increasing the pressure may decrease the flow rate.
For example, increasing $P_g$ from~$205\;\mathrm{Pa}$ to
$216\;\mathrm{Pa}$ reduces the flow rate
as the flow transitions from hex-two to hex-one.

One behavior observed near such transitions
is a bubbling oscillator~\cite{raven2006periodic}.
It is based on the interplay between the stability and
dissipation of hex-one and hex-two structures.
As the channel fills with hex-one bubbles,
the flow rate decreases due to growing dissipation.
The bubbles entering the channel become smaller
such that the hex-one structure is no longer stable.
The flow transitions to hex-two and accelerates again.
The process repeats indefinitely.
In our simulations, we obtain such an oscillator with~$L=5.3\;\mathrm{mm}$,
$T_b=0.46\,{W_0^2H}/{Q_l}$ and the inlet pressure $P_g=225\;\mathrm{Pa}$.
The flow rate plotted in~\suppreffigure{f_crystal_volume} oscillates in time.
Examples of the transitions between hex-one and hex-two
are shown in~\cref{f_crystal}.

%-% crystal.tex

%+% microfoam.tex
\subsection{Bidisperse foam generation}

A recently demonstrated microfluidic device~\cite{vecchiolla2018}
makes use of bubble-bubble pinch-off to generate bidisperse foams.
Here we reproduce its operation numerically.
The device represents a planar network $60\;\mu\mathrm{m}$ high
where a narrow channel expands with~$45^\circ$ walls to a collection channel.
Bubbles are generated periodically at an interval of~$111\;\mu\mathrm{s}$
to maintain the volume fraction of gas at~$0.645$.
At each cycle, the bubble is inserted in the narrow channel
by replacing the liquid with gas in a part of the channel.
Parameters of the simulation a summarized in \suppreftable{t_microfoam}.
Walls of the channel are no-slip boundaries.

\begin{figure}
  \centering
  \includegraphics{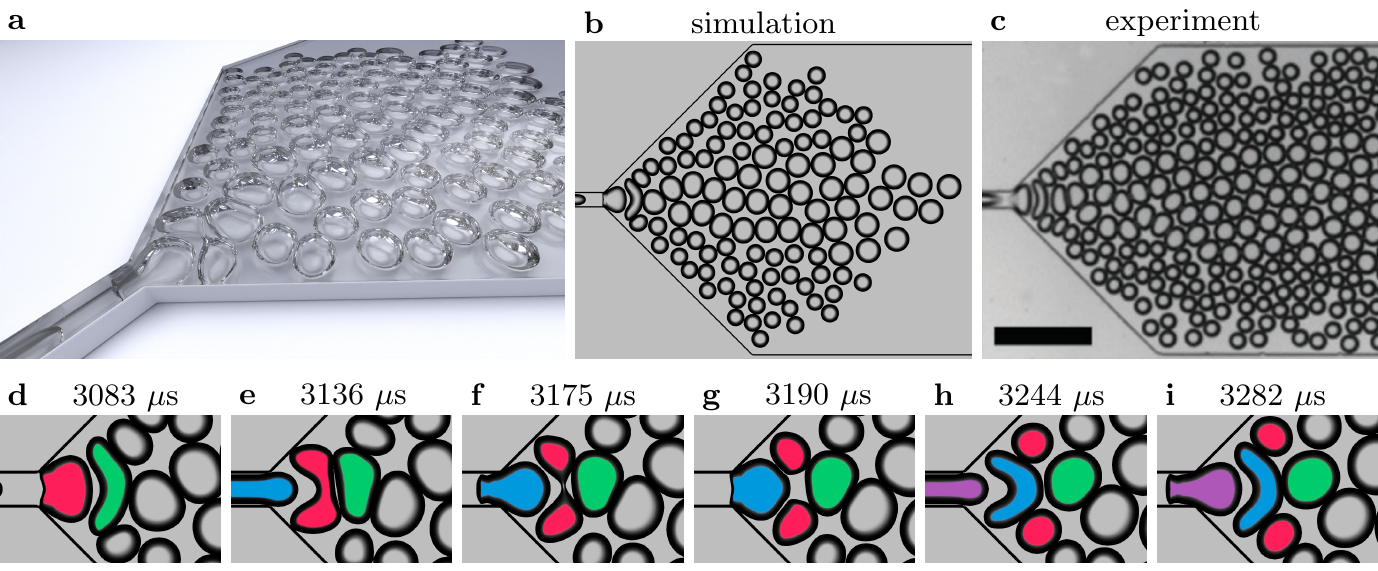}
  \caption{
    (\figbf{a})~Overall view of the microfluidic device for generation of bidisperse foams
    at~$t=8749~\mu\mathrm{s}$.
    (\figbf{b})~Snapshot from the simulation at $t=8872~\mu \mathrm{s}$.
    (\figbf{c})~Experimental image~\cite{vecchiolla2018}.  The scale bar is~$500\;\mu\mathrm{m}$.
    (\figbf{d-i}) Alternation between bubble-bubble pinch-off and elongation without breakup.
    snapshots at $t=3083, 3136, 3175, 3190, 3244$, and $3282~\mu\mathrm{s}$
    respectively. Colors highlight the same bubbles in all snapshots.
    Split bubble (red), wall bubble (green), and intact bubble (blue).
  }
  \label{f_microfoam}
\end{figure}

The overall view of the device is shown in~\cref{f_microfoam}\figbf{a}
and Movie~S6.
The snapshot from the simulation in~\cref{f_microfoam}\figbf{b} compared
to the experimental image~\cite{vecchiolla2018} (Fig. 3d therein)
indicates a good agreement with the experimental data
since both have similar shapes and positions of the split and intact bubbles.
%The histogram in~\suppreffigure{f_microfoam_hist}
%shows the distribution of the volume of the collected bubbles.

Bubbles entering the expansion alternate between two types
of behavior illustrated in~\cref{f_microfoam}\figbf{d-i}.
They either split into smaller bubbles or remain intact.
For example, one bubble (\cref{f_microfoam}\figbf{d}) enters the expansion,
elongates under the shear stress (\cref{f_microfoam}\figbf{e})
from the liquid flow and splits into two daughter bubbles (\cref{f_microfoam}\figbf{f}).
The \textit{wall bubble} confines the liquid flow.
The two daughter bubbles then migrate sideways, leaving a gap of liquid
between the wall bubble (\cref{f_microfoam}\figbf{g})
and the next incoming bubble,
which only elongates without breakup and eventually restores its shape.
Alternation of these two regimes drives the generation of bidisperse foams.

\begin{figure}
  \centering
  \includegraphics{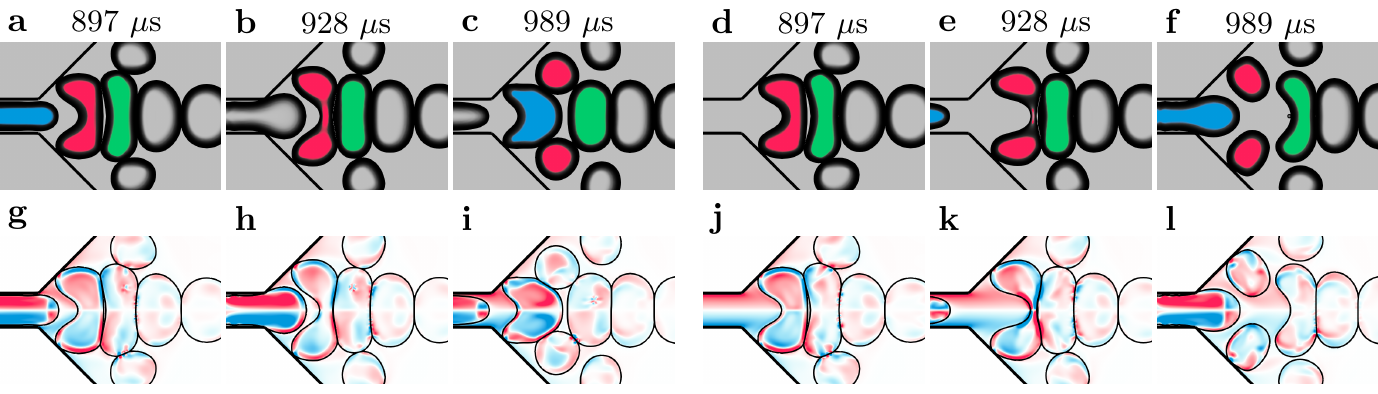}

  \caption{
    Effect of delaying the pincher bubble.
    Standard case (\figbf{a-c}, \figbf{g-i})
    and case with delayed pincher bubble (\figbf{d-f}, \figbf{j-l})
    at various time instances.
    (\figbf{a-f}) Bubble surfaces with the highlighted pincher bubble (blue),
    split bubble (red) and wall bubble (green).
    (\figbf{g-l}) Component of vorticity normal to the view plane
    (red for counterclockwise and blue for clockwise).
    Delaying the pincher bubble does not prevent breakup.
  }
  \label{f_microfoam_pinch_delay}
\end{figure}

Another part of the pinch-off process is the \textit{pincher bubble}
upstream of the split bubble
(blue in \cref{f_microfoam}\figbf{e} and
purple in \cref{f_microfoam}\figbf{h}).
One question arising here is whether the pincher bubble actually causes the breakup.
The explanation given in the experimental study~\cite{vecchiolla2018}
suggests that the pincher bubble increases the flow confinement
and the corresponding shear stresses on the split bubble, leading to breakup.
To verify this, we compare two simulations:
one with a normal pinch-off event
and one where the pincher bubble is delayed by~$60\;\mu\mathrm{s}$.
Both simulations were done with the length of the collection channel set to
$L=0\;\mu\mathrm{m}$ to reduce the computational cost,
still leaving a sufficient separation from the outlet.
\Cref{f_microfoam_pinch_delay}
shows that the pincher bubble does not trigger the pinch-off event.
Delaying the pincher bubble does not change the flow of the liquid
near the split bubble (\cref{f_microfoam_pinch_delay}\figbf{h} and
\cref{f_microfoam_pinch_delay}\figbf{k}).
This indicates that the pinch-off is triggered by
the wall bubble downstream rather than the pincher bubble upstream.

%-% microfoam.tex

%+% clustering.tex
\subsection{Clustering of bubbles}%
\label{s_clustering}

Clustering of bubbles floating on the surface of water is an example of
self-assembly known as the Cheerios effect~\cite{vella2005cheerios},
named after the observation that breakfast cereals floating in milk
often clamp together.
A bubble floating on the surface creates an elevation
attracting other bubbles due to buoyancy.
Many floating bubbles are hence attracted to each other and form clusters.

\begin{figure}[p]
  \centering
  \includegraphics{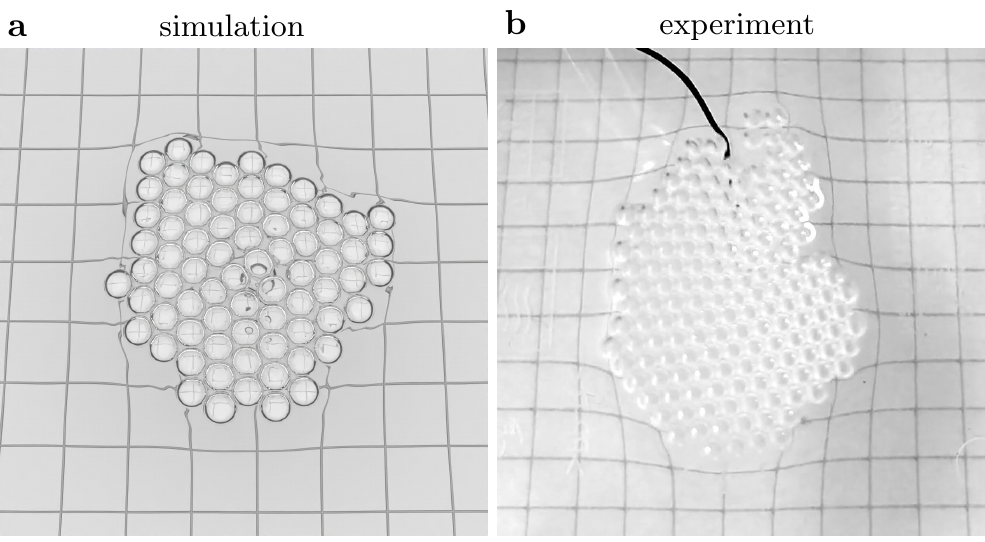}
  \caption{
    Clustering of bubbles floating on water.
    (\figbf{a})~Simulation using the Multi-VOF method
    (\figbf{b})~Experiment in soapy water. The size of square cells is $4\;\mathrm{mm}$.
  }
  \label{f_clustering}
\end{figure}

\Cref{f_clustering} shows clusters of bubbles obtained numerically
and experimentally.
In the experimental setup,
a large tank is filled halfway with tap water and a common detergent.
One end of a tube with an inner diameter of about~$0.3\;\mathrm{mm}$ is
submerged into water, and the other end is connected to a syringe filled with
air. The plunger is then abruptly pushed until bubbles start to appear.
This generates bubbles of about $2\;\mathrm{mm}$ in diameter.
To visualize the deformation of the surface, the bottom of the tank
is covered with a patterned sheet.
In the simulation, spherical bubbles are generated at the bottom
at regular intervals. Parameters are in \suppreftable{t_column}.
Both in the experiment and the simulation, bubbles floating on the surface form
clusters and organize in a hexagonal lattice.
Movie~S7 shows the simulation results.

%-% clustering.tex

%+% waterfall.tex
\subsection{Foaming waterfall}%
\label{s:waterfall}

\begin{figure}
  \centering
  \includegraphics{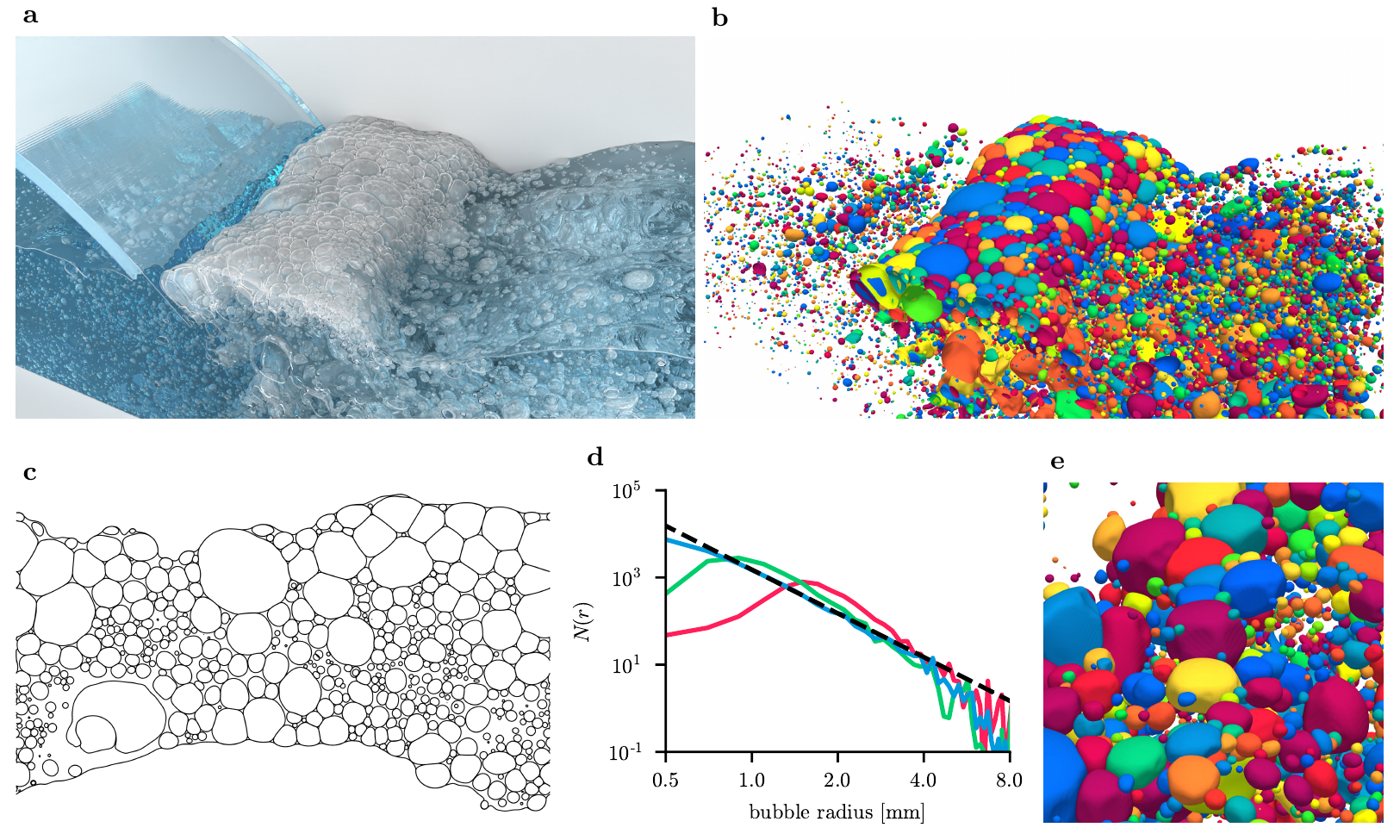}

  \caption{
    Foaming waterfall.
    (\figbf{a})~Overall view of the interface at
    $t=1.2\;\mathrm{s}$.
    (\figbf{b})~Same view showing bubbles under the surface with arbitrary colors.
    (\figbf{c})~Horizontal cross-section of the interface at
    $t=1.2\;\mathrm{s}$.
    Clusters of bubbles on the surface show
    characteristic features of foam:
    bubbles are separated by thin membranes (lamellae)
    with multiple junctions (Plateau borders).
    (\figbf{d})~%
    Histogram of the bubble radius.
    $N(r)$ is the number of bubbles
    with the equivalent radius in the range $r\pm0.1\;\text{mm}$.
    Results for different mesh sizes of
    $96$~\key{110}, $192$~\key{120}, and $384$~\key{130} cells in the height
    compared to scaling law~\cite{garrett2000}
    $N(r)\propto r^{-10/3}$~\key{200}.
    The values are averaged in time over~$1<t<1.2\;\mathrm{s}$.
    (\figbf{e})~Close-up of (\figbf{b}) with half of the bubbles removed.
  }
  \label{f_waterfall}
\end{figure}

Natural surfactants in sea water can suppress coalescence as well.
Oceans are covered with foam generated by breaking waves.
The following application is an example of such flows
in the limiting case without coalescence.
A rectangular tank $100\;\mathrm{mm}$ high is filled halfway
with water. A waterfall enters the tank at a given velocity.
\suppreftable{t_waterfall} lists the simulation parameters.
Results of the simulation are in \cref{f_waterfall}
with additional snapshots in \suppreffigure{f_waterfall_series} and Movie~S8.
On the finest mesh consisting of $768\times384\times384$ cells,
the simulation took 20 hours on 1152 compute nodes of the Piz Daint supercomputer
equipped with 12-core CPU Intel Xeon E5-2690 v3 processors.
Two major mechanisms of air entrainment~\cite{kiger2012}
are observed in this simulation:
entrapment of a tube of air when the sheet of water impacts the surface
and entrainment around the impact site as the waterfall
drags air into the water.
The entrained bubbles rise to the surface and create a layer of foam.
As seen from the horizontal cross-section of the foam in~\cref{f_waterfall},
the bubbles are separated by thin membranes (lamellae)
that form multiple junctions (Plateau borders)
at angles approaching~$120^\circ$.

The distribution of the bubble size in~\cref{f_waterfall}
matches a scaling law~\cite{garrett2000}~$N(r)\propto r^{-10/3}$,
where~$N(r)$ is the number of bubbles of radii in the range $r\pm0.1\;\mathrm{mm}$.
The model stems from the assumption that
the inflow of air per unit volume is constant,
and the number of bubbles depends only on the turbulent dissipation rate
and the bubble radius.
This scaling law is commonly observed for bubbles generated by breaking waves
and has been reported in experimental~\cite{deane2002}
and numerical~\cite{deike2016} studies.

%-% waterfall.tex

%+% discussion.tex
\section{Discussion}
\label{s_discussion}

The multilayer volume-of-fluid method
can simulate flows with many bubbles and drops that do not coalesce.
It represents many bubbles with a fixed number of
volume fraction fields and assigns colors to bubbles to distinguish them.
An additional technique of interface regularization
based on forward-backward advection
improves the accuracy of the advection scheme.
The presented applications show that the method can reproduce experiments
on generation of foam in microfluidic devices
and clustering of bubbles floating in water.

The proposed methodology advances the state of the art in simulations of flows with multiple interfaces in  the following categories:
\begin{enumerate}

  \item Efficiency:
    The method uses a fixed number of volume fraction fields on an Eulerian mesh.
    The computational complexity of the advection algorithm is linear with the number of cells
    and does not depend on the number of bubbles.

  \item Compatibility with existing methods:
    The method is compatible with  existing stencil-based methods
    for interface capturing and curvature estimation,
    including the popular volume-of-fluid and level-set methods.For dry foams, the method can recover the results of~VIIM at
comparable resolutions.

  \item Capturing topology changes  and multiple bubble junctions:  Breakup and coalescence of
    bubbles including thin liquid films and triple lines are captured without employing ad-hoc parameters.

  \item Coupling with models of film drainage and rupture:
    Coalescence can be controlled by assigning
    appropriate color functions to interacting bubbles.

    \item High performance implementation:
      The method only involves stencil operations and is readily integrated in high performance software for structured grids.
\end{enumerate}

We believe that Multi-VOF opens new horizons for
simulating a wide variety of flows from the micro to the macroscale,
including  wet foams, turbulent flows with bubbles, suspensions and emulsions in microfluidics.  Moreover, the efficiency of the code allows for extensive studies in control and optimization of bubbly flows. 

\subsection{Limitations}

The method describes the complete prevention of coalescence and a
an empirical  criterion  is used in cases where the residence time of bubbles is finite.
Since bubbles are distinguished as connected components
of the volume fraction field,
the method does not prevent coalescence if a deformed bubble folds back onto itself. Two small bubbles can penetrate each other and form concentric configurations when their  radius is comparable to one computational cell.
The method can be improved by adopting 
a semi-implicit discretization of the surface tension
force~\cite{cottet2016semi}.

Generation of  bubbles requires additional modeling.
In simulations, we forcibly separate the air thread at regular intervals.
Otherwise, the gas thread remains continuous unless the inlet pressure is sufficiently low.
Such continuous regimes are observed experimentally
under certain conditions~\cite{anna2003}
but not in the experimental study of interest~\cite{raven2009}.
One explanation for this discrepancy is
the effect of wetting~\cite{anna2003} that
may narrow the gap between the liquid-gas interface
and the channel walls since the static contact angle of polydimethylsiloxane (PDMS) is
$104^\circ$~\cite{mata2005} while our model assumes~$180^\circ$.
Another possibility is that the viscous flow of the liquid upstream of the
orifice is not sufficiently resolved in the simulations.
%-% discussion.tex

%+% methods.tex
\section{Methods}
\label{s_methods}
\subsection{Multilayer fields}
\label{s_multilayer}

\begin{figure}
  \centering
  \includegraphics{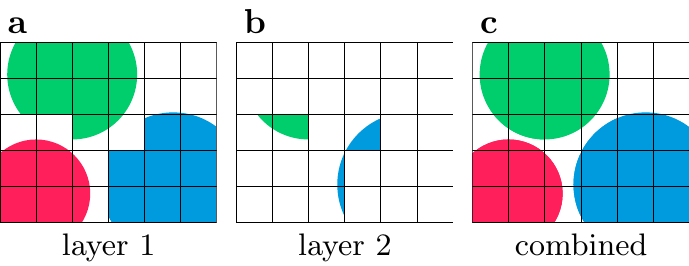}

  \caption{Two-layer volume fraction field representing three bubbles.
    Individual layers (\figbf{a,b}) contain at most one bubble per cell,
    while the combined field (\figbf{c}) describes overlapping interfaces.
  }
  \label{f_sketch_layers}
\end{figure}

Consider a discrete domain~$\Omega$ consisting of
cells~$c\in\Omega$, where the number of cells is $|\Omega|=N_\mathrm{cells}$.
A conventional cell field~$\phi\colon c\mapsto \phi_c$
is a mapping from cell~$c$ to a value.
A \textit{cell-color} field~$\hat\phi\colon(c,q)\mapsto \hat\phi_c(q)$ is a
mapping from cell~$c$ and \textit{color}~$q\in\mathbb{R}$ to a value.
Overlapping bubbles can be represented by a single cell-color field
if each bubble is assigned a unique color.
The restriction operation $\big.\hat\phi\big|_q$ constructs
a conventional field $\big.\hat\phi\big|_q\colon c \mapsto \hat\phi_c(q)$
from a cell-color field~$\hat\phi$ given a color $q$.
Using this operation, any standard routine,
such as computing the normals or solving the advection
equation, can be applied to a cell-color field by individually
selecting all possible colors.
To store a cell-color field~$\hat\phi$,
we use a sequence of conventional fields for values and colors separately.
Assume that any cell contains at most~$L$ bubbles
and their shapes are represented by the cell-color volume fraction
field~$\hat\alpha$.
The colors are stored in fields $q^l:c\mapsto q_c^l,\;l=1,\dots,L$
defined in each cell~$c$ as
\begin{equation*}
  q_c^l =
  \begin{cases}
    q_l, &1\leq l \leq L'\,, \\
    q_\mathrm{none}, &\text{otherwise}\,,
  \end{cases}
\end{equation*}
where~$\{q_1, \dots, q_{L'\leq L}\}$ are all colors for
which~$\hat\alpha_c(q)>0$
and $q_\mathrm{none}$ is a distinguished \textit{none} color
(e.g. $q_\mathrm{none}\coloneqq -1$).
The corresponding
values are stored in fields $\phi^l:c\mapsto \phi_c^l,\;l=1,\dots,L$
\begin{equation*}
  \phi^l = \begin{cases}
    \hat\phi(q_c^l), &q_c^l \neq q_\mathrm{none}\,, \\
    \textit{undefined}, &\text{otherwise}\,.
  \end{cases}
\end{equation*}
The pairs $(\phi^l, q^l)$ are referred to as \textit{layers}
and the sequences $(\phi^1,\dots, \phi^L)$ and $(q^1,\dots, q^L)$
constitute a \textit{multilayer} field.
The order in which the colors are stored is insignificant,
i.e. all sequences $(q_c^1,\dots, q_c^L)$ and $(\phi_c^1,\dots, \phi_c^L)$
are equivalent up to mutual permutation.
\Cref{f_sketch_layers} illustrates two layers that describe three
overlapping bubbles.

\subsection{Advection}
\label{s_advection}

By constructing conventional fields from a cell-color field,
we can apply standard stencil-based algorithms to cell-color fields.
One such algorithm, the PLIC (Piecewise Linear Interface Characterization)
method~\cite{youngs1982} for advection, is described in the following.
The PLIC method solves the advection equation given a velocity field~$\mb{u}$
$\frac{\pd \alpha}{\pd t} + (\mb u \cdot \nabla)\, \alpha = 0.$
As the name stands, it performs a piecewise linear reconstruction
by replacing the interface in each cell with a plane.
Apart from the volume fraction field, it involves normals and plane constants.
The normals~$\mb{n}:\Omega\rightarrow\mathbb{R}^3$ are estimated from the
volume fractions using the mixed Youngs-centered scheme~\cite{aulisa2007}
and the plane constants are computed from the normals and volume fractions
using explicit formulas~\cite{scardovelli2000}.
The fluid volume is reconstructed in each cell with a polyhedron
formed by cutting the cell with the plane.
The fluxes are then computed by advecting the polyhedrons according to the given
velocity~\cite{aulisa2007,weymouth2010}.
The discretization uses directional splitting,
and a step in one direction can be schematically written
in terms of discrete operators~$\mathcal{N}$ and $\mathcal{A}$:
$\mb{n}_c = \mathcal{N}\big(S^c[\alpha]\big)$ and
$\alpha^{\text{new}}_c = \mathcal{A}\big(S^c[\alpha], S^c[\mb{n}]\big)$,
where~$S^c=\big(c_1, \dots, c_{27} \big)$~is the sequence of cells in the $3\times3\times3$
stencil centered at $c$ and
$S^c[\phi]=\big(\phi_{c_1}, \dots, \phi_{c_{27}})$
are the corresponding values of a field~$\phi$.
To apply this method to a cell-color field, the same procedure is repeated for all colors:
$\hat{\mb{n}}_c(q) = \mathcal{N}\big(S^c[\left.\hat\alpha\right|_q]\big)$ and
$\hat\alpha^{\text{new}}_c(q) = \mathcal{A}\big(S^c[\left.\hat\alpha\right|_q], S^c[\left.\hat{\mb{n}}\right|_q]\big)$.
In terms of multilayer fields~$q^l$, $\alpha^l$ and $\mb{n}^l$,
the normals are computed with the following algorithm
\begin{center}
\begin{minipage}{12cm}
\begin{algorithmic}[H]
\For{$c\in\Omega$\;,\; $l=1,\dots,L$}
  \If{$q_c^l \neq q_\mathrm{none}$}
    \For{$i=1,\dots,27$}
      \State $c' \leftarrow S^c_i$
      \State $\bar\alpha_i \leftarrow 0$
      \For{$l'=1,\dots,L$}
        \If{$q_{c'}^{l'} = q_c^l$}
          \State $\bar\alpha_i \leftarrow \alpha_{c'}^{l'}$
        \EndIf
      \EndFor
    \EndFor
    \State $\mb{n}_c^l \leftarrow \mathcal{N}(\bar\alpha)$
  \EndIf
\EndFor
\end{algorithmic}
\end{minipage}
\end{center}
The total number of operations of this algorithm
is~$\mathcal{O}(L^2\,N_\mathrm{cells})$.
Before proceeding with advection, the normals are corrected.
When two or more interfaces enter one cell, we ensure that their normals
are parallel.
From the estimated normals~$(\mb{n}_1,\dots,\mb{n}_{L'})$
we compute the average
$\mb{n}_\mathrm{avg} = \frac{1}{L'}\sum_{l=1}^{L'}\mb{n}_l \,\sign(\mb{n}_l \cdot \mb{n}_1)$
and overwrite the normals as
$\mb{n}_l \leftarrow \mb{n}_\mathrm{avg} \sign(\mb{n}_l \cdot \mb{n}_\mathrm{avg})$.
This correction prevents mutual penetration of interfaces.

The advection step is done similarly,
but includes also new colors found in upwind cells.
If the new volume fraction for an upwind cell color is positive,
the color is added to the current cell.
Colors corresponding to zero volume fractions are removed.
For the number of layers we find sufficient~$L=4$
based on the case of close packing of rising bubbles in~\suppreffigure{f_packing_rise}.

\subsection{Connected-component labeling}

Prevention of coalescence requires that all bubbles have unique colors.
These unique colors can be assigned from initial conditions.
For instance, using the indices of bubbles as colors.
If the set of bubbles remains the same,
the colors remain unique throughout the simulation.
However, new colors are needed for injected bubbles or bubbles formed
during breakups.
To detect breakups and assign unique colors to all bubbles,
we use connected-component labeling.

Starting with the old color fields~$q^l$ and volume fraction
fields~$\alpha^l$,
we construct new color fields~$\tilde{q}^l$
in which all bubbles have unique colors.
Different bubbles are identified as connected components in the volume fraction
field.
Two neighboring cells and layers $(c,l)$ and $(c',l')$
are connected with an edge if they have positive volume fractions
$\alpha^l_c>0$ and $\alpha^{l'}_{c'}>0$ and equal colors $q^l_c=q^{l'}_{c'}$.

To detect the connected components,
we first initialize~$\tilde{q}_c^l$ with unique colors for all cells and layers.
For instance, using an integer index enumerating all cells and layers
(in total, $N_\text{cells}\,L$ colors).
Then we iterate until convergence over all pairs of connected cells and layers
choosing the minimal color.
The procedure is implemented by the following algorithm
\begin{center}
\begin{minipage}{12cm}
\begin{algorithmic}[H]
  \State $i \leftarrow 0$
  \For{$c\in\Omega$\;,\; $l=1,\dots,L$} \Comment{initialize with unique colors}
    \State $\tilde{q}_c^l \leftarrow i$
    \State $i \leftarrow i+1$
  \EndFor

  \Repeat \Comment{propagate smaller colors until convergence}
    \State $\textit{converged}\leftarrow \True$
    \For{$c\in\Omega\;,\;l=1,\dots,L$}
      \If{$q_c^l \neq q_\mathrm{none} \And \alpha_c^l > 0$}
        \For{$c' \in S^c\;,\;l'=1,\dots,L$}
          \If{$q_{c'}^{l'} = q_c^l
              \And \alpha_{c'}^{l'} > 0
              \And \tilde{q}_{c'}^{c'} < \tilde{q}_{c}^{l}$}
            \State $\tilde{q}_c^l \leftarrow \tilde{q}_{c'}^{l'}$
            \State $\textit{converged}\leftarrow \False$
          \EndIf
        \EndFor
      \EndIf
    \EndFor
  \Until{\textit{converged}}
\end{algorithmic}
\end{minipage}
\end{center}
\suppreffigure{f_labeling} illustrates the algorithm on a case
with one layer and three connected components.

\subsection{Two-component incompressible flows}
\label{s_flow}

The model of two-component incompressible flows consists of
Navier-Stokes equations for the mixture velocity~$\mb u$ and pressure~$p$
\begin{align*}
  \nabla \cdot \mb u &= 0\,, \\
  \rho\Big(\frac{\pd \mb u}{\pd t} + (\mb u \cdot \nabla)\, \mb u\Big)
  &=
  -\nabla p + \nabla \cdot \mu (\nabla \mb u + \nabla \mb u^T )
  + \mb f_\sigma + \rho\mb g
\end{align*}
and the advection equation for the volume fraction~$\alpha$
\begin{equation*}
  \frac{\pd \alpha}{\pd t} + (\mb u \cdot \nabla)\, \alpha = 0\,,
\end{equation*}
with the mixture density~$\rho=(1-\alpha)\rho_1 + \alpha\rho_2$,
dynamic viscosity~$\mu=(1-\alpha)\mu_1 + \alpha\mu_2$,
surface tension force $\mb f_\sigma = \sigma \kappa \nabla \alpha$
and gravitational acceleration~$\mb g$,
where $\sigma$ is the surface tension and $\kappa$ is the interface curvature.
The mixture flow equations are discretized
with the projection method~\cite{bell1989},
and the advection equation is solved 
using the procedure described in the~\nameref{s_advection}.
The mixture density and viscosity fields are computed
from the combined volume fraction field
$\alpha = \min\big(1, \sum_q \left.\alpha\right|_q \big)$.
The surface tension force is computed by summation over all colors
$\mb{f}^\sigma = \sum_q \sigma \left.\kappa\right|_q \left.\nabla \alpha\right|_q$
and the curvature~$\left.\kappa\right|_q$ is estimated
from the volume fractions using the method of particles~\cite{partstr}.
We apply a technique described in the \suppnameref{s_methods} to regularize the interface
produced by the advection scheme which does not affect its asymptotic
convergence or conservation properties
but results in smoother surfaces at low resolutions.
To simulate flows in complex geometries on a Cartesian mesh,
we employ the method of embedded boundaries~\cite{colella2006}
which approximates the shape of the body with cut cells.

\subsection{Code availability}

The simulations are performed using the open-source solver Aphros
available at \url{https://github.com/cselab/aphros}.
The configuration files are located in
\url{https://github.com/cselab/aphros/tree/master/examples/205_multivof}.
An online demonstration of the method in two dimensions
is at \url{https://cselab.github.io/aphros/wasm/hydro.html}.

%-% methods.tex

\bibliography{main}

\end{document}

\section{Supplementary Movies}

\url{https://polybox.ethz.ch/index.php/s/wtFIhSd77RG3Nyr}

\begin{itemize}
  \item 
  \verb`s1_hex_one.mp4`~--- microfluidic crystal, hex-one
  \item 
  \verb`s2_hex_two.mp4`~--- microfluidic crystal, hex-two
  \item 
  \verb`s3_hex_three.mp4`~--- microfluidic crystal, hex-three
  \item 
  \verb`s4_hex_four.mp4`~--- microfluidic crystal, hex-four
  \item 
  \verb`s5_oscillator.mp4`~--- microfluidic crystal, oscillator
  \item 
  \verb`s6_bidisperse.mp4`~--- bidisperse foam
  \item 
  \verb`s7_clustering.mp4`~--- clustering of bubbles floating in water
  \item 
  \verb`s8_waterfall.mp4`~--- foaming waterfall
\end{itemize}

\section{Supplementary Methods}
\label{s_methods}

\subsection{Connected component labeling (Supplementary)}

If the domain is decomposed into rectangular blocks for distributed parallelization,
the algorithm is first applied to each block,
then the halo cells are exchanged through communication
and the whole procedure is repeated until global convergence.
The total number of communication steps
is the maximum length of a connected component divided by the block size.
To reduce the number of communication steps, we use a heuristic
for faster propagation of colors over the domain.
The heuristic is executed before each iteration of the main algorithm.
For each block, we consider only one corner cell $(c,l)$
and its neighbors $(c',l')$ and, if they are connected,
collect the corresponding
pairs of colors~$\tilde{q}_{c}^{l}$ and $\tilde{q}_{c'}^{l'}$
and transfer them all to one processor.
Then we perform connected component labeling on this set of pairs
and find a set of connected components.
Finally, we broadcast them to other processors
and replace each color with the minimal element of its connected component.
With this heuristic, large connected components are identified
after fewer iterations.

\begin{figure}[H]
  \centering
  \includegraphics{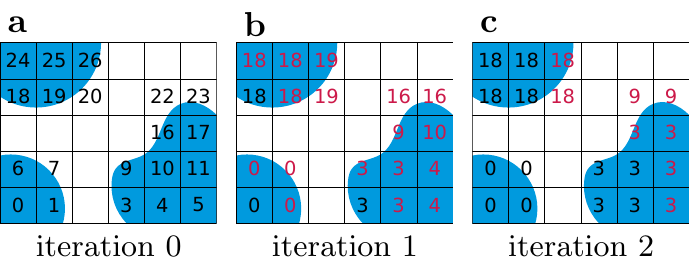}

  \caption{Example of connected component labeling on a $6\times5$ grid
  with $L=1$.
  Current colors~$\tilde{q}_c^l$ are shown with numbers.
  Initial colors (\text{iteration 0}) are indices enumerating all cells.
  Each iteration selects the minimal color over the $3\times 3$ stencil.
  Colors updated at the current iteration are highlighted in red.
  }
  \label{f_labeling}
\end{figure}

\subsection{Interface regularization}
\label{s_sharp}

Volume-of-fluid methods based on geometrical reconstruction,
such as PLIC \cite{youngs1982} employed in this work,
offer good approximation properties, ensure boundness and conservation.
They sharpen the interface of advected objects and hence reduce the numerical
diffusion.
However, the sharpening effect depends on the advection velocity.
The choice of the frame of reference and the location of advected
objects in a nonuniform velocity field affect the solution, which is undesirable.
To mitigate this and also improve the accuracy at lower resolutions,
we propose a technique for interface regularization based on PLIC.
The idea is to apply forward and backward advection with a uniform velocity.
This procedure is supposed to reduce the thickness of the interface
and remove small spurious interface fragments.
At the same time, it maintains volume conservation and boundness
and does not affect the asymptotic convergence properties of the method.

\begin{figure}[H]
  \centering
  \includegraphics{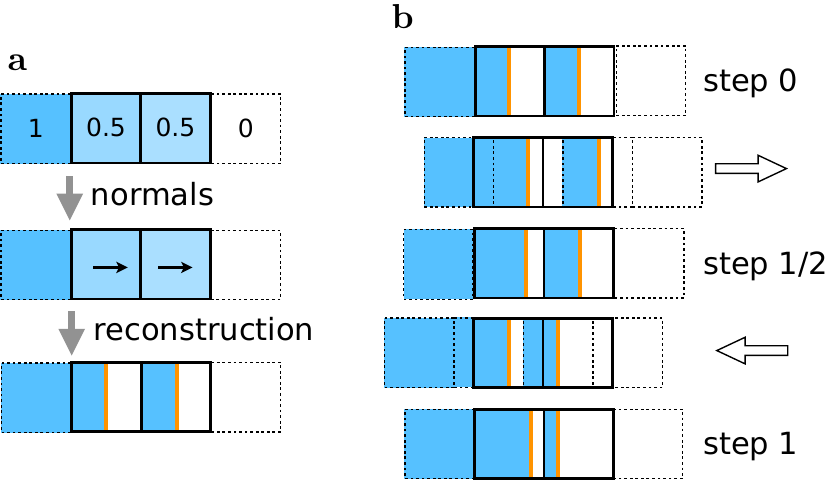}
  \caption{
    (\figbf{a})
    Stages of PLIC reconstruction in one dimension.
    The interface in each cell is replaced by a plane (orange)
    which is oriented by the estimated normal
    and cuts the cell at the given volume fraction.
    (\figbf{b})
    Forward-backward advection reducing the interface thickness.
  }
  \label{f_sharp_sketch}
\end{figure}

In one dimension, the procedure consists of two advection sweeps
illustrated in~\suppreffigure{f_sharp_sketch}.
The initial interface~(step $1$) spans two cells with volume fractions
$(0.5,0.5)$.  The forward step moves the interface to the right (step $1/2$)
and the volume fractions take values~$(0.8, 0.5)$ as one cell fills with liquid
from upstream.  The backward step moves the interface to the left (step $1$)
reducing the volume fraction in the other cell,
so they reach values~$(0.8, 0.2)$.
Note that after two steps, the field is again symmetric.
In three dimensions, there are six advection sweeps
with uniform velocities
$\mb{u}= (U,0,0)$, $(-U,0,0)$, $(0,U,0)$, $(0,-U,0)$, $(0,0,U)$, and $(0,0,-U)$.
They are repeated after every time step of the simulation.
The only parameter of the regularization is the advection velocity
defined by the CFL number, which we set
to $U\Delta t/h=0.1$ for all simulations.
To reduce the computational cost, the pairs of sweeps
in each direction can be spread among every three consecutive time steps
leaving only two sweeps per time step.

\begin{figure}[H]
  \centering
  \includegraphics{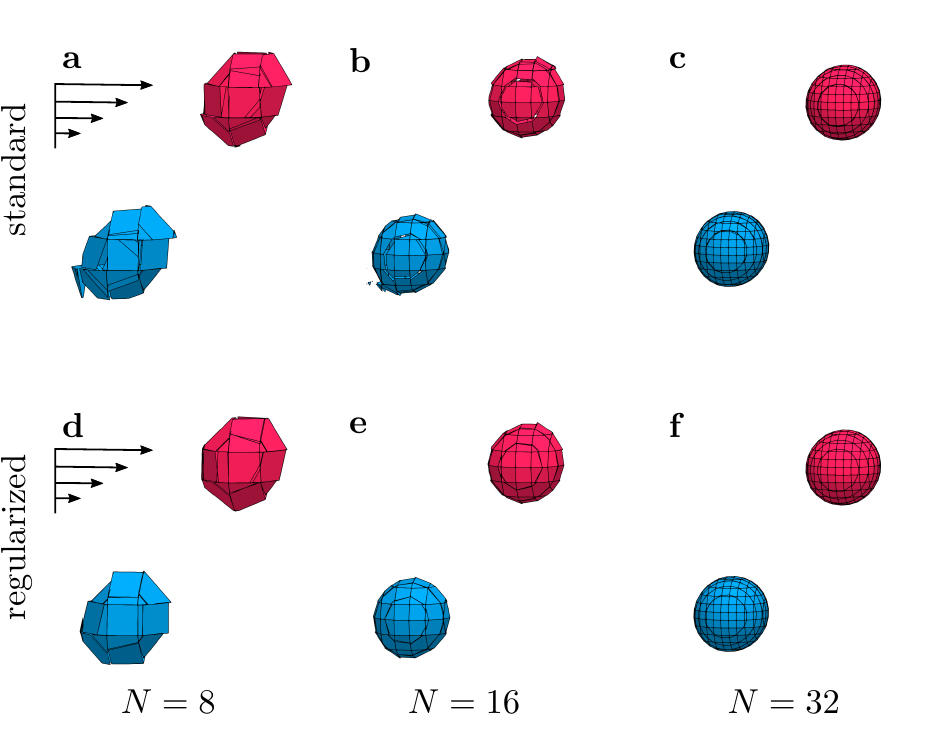}

  \caption{
    Drops in Couette flow with the standard advection~(\figbf{a-c})
    and with interface regularization~(\figbf{d-f}) on a mesh of $N^3$ cells.
    Snapshots at $t=0.75$.
  }
  \label{f_sharp_couette}
\end{figure}

As noted above, the solution by the standard PLIC method
depends on the choice of frame of reference.
Moreover, at low resolutions the method can produce
small spurious interface fragments.
These effects are evident from the evolution of drops
in a Couette flow shown in~\suppreffigure{f_sharp_couette}.
The domain is a unit cube periodic in two directions and no-slip walls at
the bottom and top with imposed velocities of 0 and 1 respectively.
The mesh size is $N^3$ with $N=8,16$ or $32$.
Initially, the drops are spheres of radius~$0.15$.
The Reynolds number is $\mathrm{Re}=\rho/\mu=100$,
the capillary number is $\mathrm{Ca}=\mu/\sigma=0.1$,
and both components have the same density and viscosity.
Since the flow is symmetric, in the exact solution the drops
need to be symmetric as well.
However, the numerical solution lacks this symmetry since the drops move
at different speeds.
Solving the problem with the standard method without regularization,
we observe that the top (red) and bottom (blue) drops have different shapes
and small interface fragments appear
near the bottom drop at lower resolutions of $N=8$ and $16$.
Enabling the regularization not only removes the spurious fragments
but also makes the shapes of the two drops more similar
thus reducing their dependence on the frame of reference.

\begin{figure}[H]
  \centering
  \includegraphics{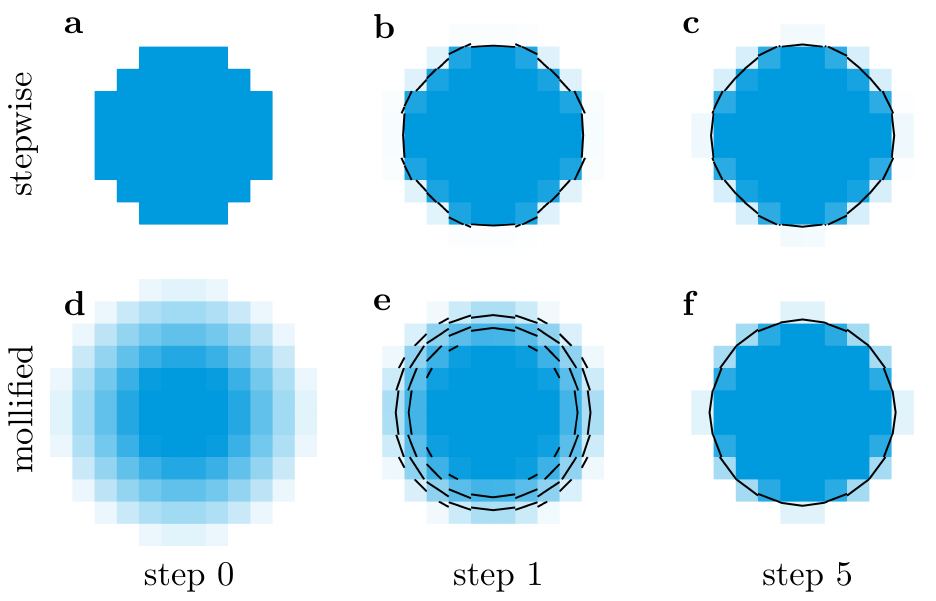}

  \caption{
    Interface regularization applied to
    two fields representing a circle:
    stepwise~(\figbf{a-c}) and radial piecewise linear~(\figbf{d-f}).
  }
  \label{f_sharp_circle}
\end{figure}

Another possible application of this technique is the regularization of initial
conditions.
\suppreffigure{f_sharp_circle} demonstrates two different fields
representing a circle: a stepwise approximation and a mollified field.
In both cases, a few steps of the regularization produce
an interface with a thickness of one cell.

\section{Supplementary Note: Verification}
\subsection{Time-reversed advection}

\begin{figure}[H]
  \centering
  \includegraphics{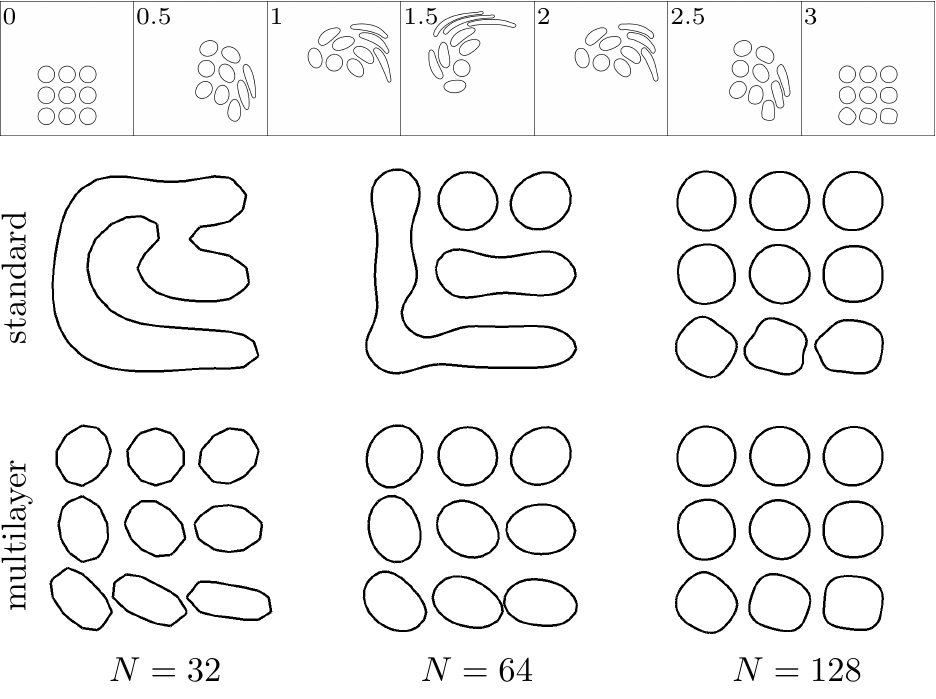}

  \caption{
    Time-reversed advection of circles.
    Snapshots at times $t=0$, $0.5$, $1$, $1.5$, $2$, $2.5$, and $3$ (top)
    on the finest mesh of $64^2$ cells with multilayer advection.
    Final profiles at $t=3$ with
    standard advection (middle) and multilayer advection (bottom)
    on a sequence of meshes of~$N^2$ cells with $N=32$, $64$ and $128$.
    The exact solution at~$t=3$ coincides with the initial profile.
  }
  \label{f_timereversed}
\end{figure}

The following setup is based on a classical test case for advection schemes.
The initial volume fraction field describes nine circles
and the velocity field defined in a two-dimensional
unit domain represents a single vortex reversed at time~$t=1.5$
\begin{gather}
  \mb{u}(x, y, t) = 
  \begin{cases}
    (\sin{\pi x}\, \cos{\pi y},\, -\cos{\pi x}\, \sin{\pi y}), &t < 1.5\,,\\
    (-\sin{\pi x}\, \cos{\pi y},\, \cos{\pi x}\, \sin{\pi y}), &t \geq 1.5\,.
  \end{cases}
\end{gather}
Since the advection equation is time-reversible,
the exact solution at~$t=3$ coincides with the initial profile.
\suppreffigure{f_timereversed} shows the solution in time
and also compares the final profiles obtained by the standard PLIC advection
and our multilayer method on a mesh of $N^2$ cells for $N=32$, $64$ and $128$.
As seen from the final profiles at lower resolutions of $N=32$ and $64$,
the standard advection method merges some of the circles into larger
objects.  This effect is the origin of numerical coalescence.
The deformations reduce that distances between the interfaces,
and eventually they merge since the method cannot describe interfaces
overlapping in the same cell.
On the other hand, our multilayer advection method
correctly maintains all circles separated.
Both methods converge to the same solution with mesh refinement.

\subsection{Buoyancy-driven breakup}

\begin{figure}[H]
  \centering
  \includegraphics{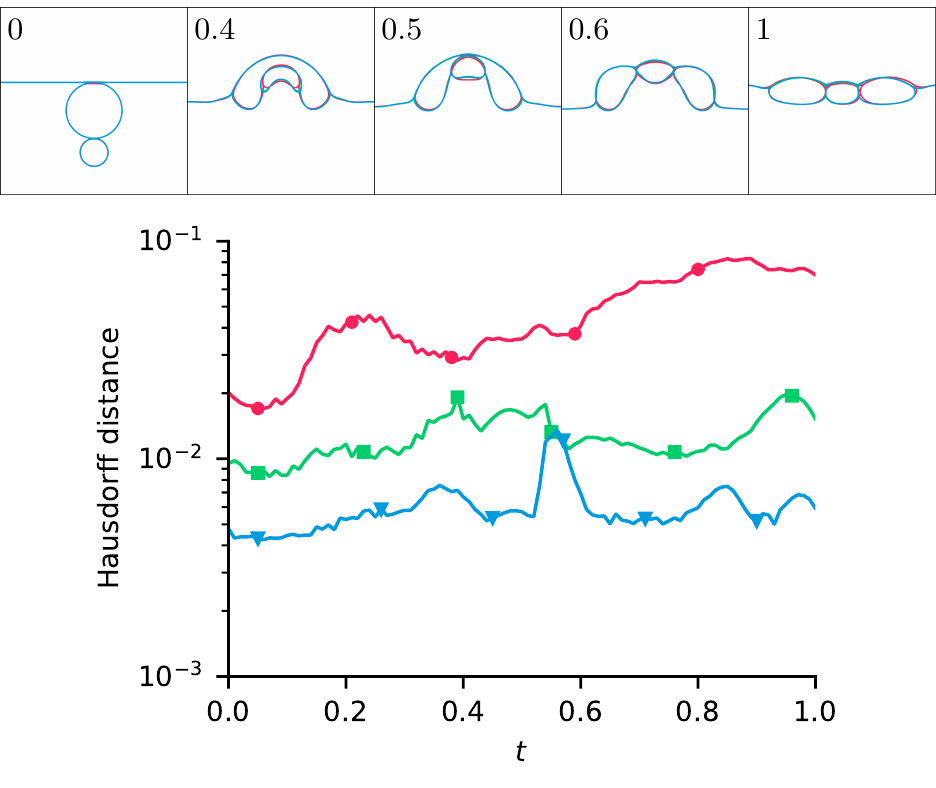}

  \caption{
    Buoyancy-driven breakup of bubbles.
    Snapshots at~$t=0$, $0.4$, $0.5$, $0.6$, and $1$ (top).
    Hausdorff distance between the interfaces
    computed on meshes with~$(N/2)^2$ and $N^2$ cells (bottom).
    Lines correspond to
    $N=64$~\key{111}, $N=128$~\key{122}, and $N=256$~\key{133}.
  }
  \label{f_risingconv}
\end{figure}

This case shows a buoyancy-driven breakup of two bubbles on a free surface
and serves as a convergence test.
The problem is solved in the unit domain $[0,1]^2$
bounded by free-slip walls in the vertical direction
and periodic in the horizontal direction.
The mesh consists of $N^2$ cells for $N=32$, $64$, $128$, and $256$.
The initial velocity is zero, and the volume fraction field represents a layer
of gas $0.6<y<1$, one bubble of radius 0.15 and one bubble of radius 0.075.
Parameters of the problem are the density~$\rho_1=1$ and~$\rho_2=0.01$,
viscosity~$\mu_1=0.01$ and~$\mu_2=0.0001$, gravitational acceleration $g=5$
and surface tension~$\sigma=0.01$.
\suppreffigure{f_risingconv} presents the solution in time for various~$N$
and the corresponding Hausdorff distance between the interfaces
computed on meshes with~$(N/2)^2$ and $N^2$ cells.
The bubbles rise and deform the free surface.
The small bubble underneath penetrates the larger bubble and splits it into two
equal parts. In the final configuration, three bubbles rest on the surface.
The Hausdorff distance at $t=0.4$ computed for $N=64$, $128$ and $256$
amounts respectively to $0.0291$, $0.0152$ and $0.0067$,
which indicates a first order convergence rate.

\subsection{Packing of rising bubbles}
\label{s_rise}

The case of close packing of rising bubbles demonstrates
how overlapping interfaces are distributed over multiple layers.
From this problem, we determine the number of layers~$L$ to use in other
simulations and initially set it to~$L=8$.
The problem is solved in the unit domain $[0,1]^3$
on a mesh of $64^3$ cells bounded by free-slip walls in the vertical direction
and periodic in the other directions.
The initial velocity is zero, and the volume fraction field represents a layer
of gas $0.8<y<1$ together with 397 bubbles of radius 0.05 placed uniformly over
the remaining volume.
Parameters of the problem are the density~$\rho_1=1$ and~$\rho_2=0.01$,
viscosity~$\mu_1=0.01$ and~$\mu_2=0.0001$, gravitational acceleration $g=5$
and surface tension~$\sigma=0.1$.

\begin{figure}[H]
  \centering
  \includegraphics{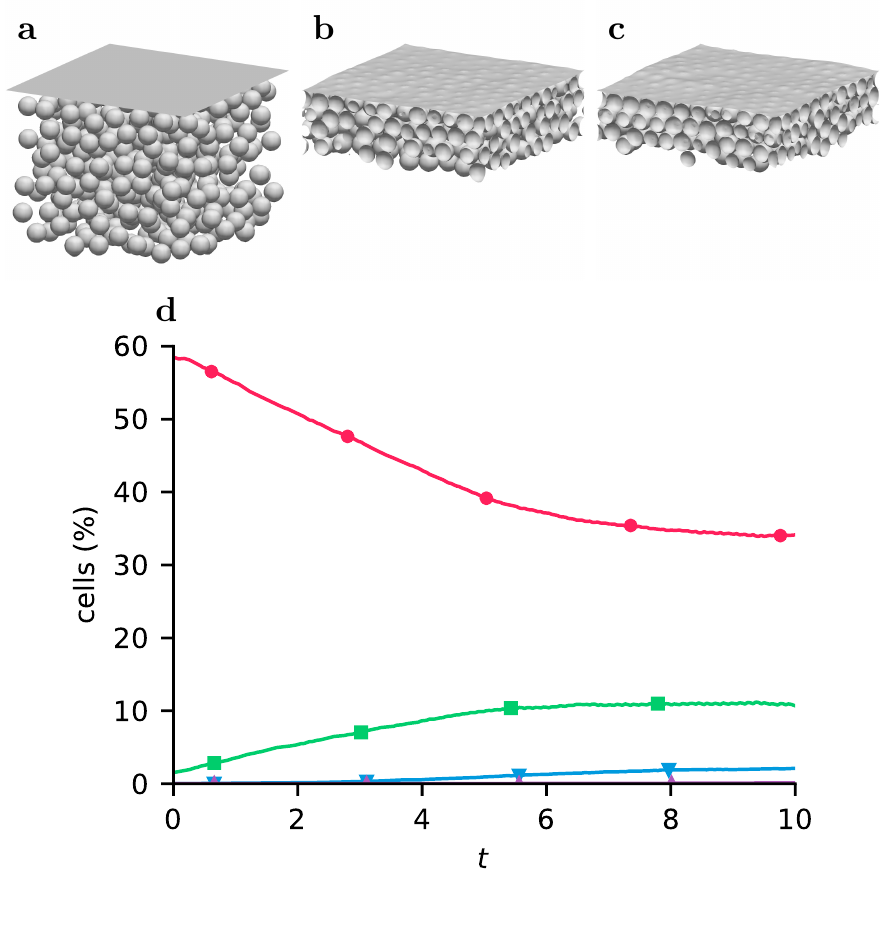}
  \caption{
    Packing of rising bubbles.
    (\figbf{a-c})~Snapshots of the interfaces at $t=0,\;5$ and~$10$.
    (\figbf{d})~Percentage of cells with
    one~\key{111}, two~\key{122}, three~\key{133}, and four~\key{144} interfaces.
  }
  \label{f_packing_rise}
\end{figure}

Snapshots of the bubbles are shown in~\suppreffigure{f_packing_rise}.
The bubbles rise and closely pack creating bulges on the free surface.
\suppreffigure{f_packing_rise} shows the fraction of cells
depending on the number of interfaces they contain.
The initial separation between the bubbles is sufficiently large
so that most cells contain only one interface.
As the bubbles rise, the gaps between them reduce
and fraction of cells with multiple interfaces increases.
At the final time~$t=10$, the percentages of cells containing from one to
six interfaces amount respectively to
$34.11$, $10.75$, $2.13$, $0.13$, $0.0034$, and $0.00076\%$.
None of the cells contain more than six interfaces.
Since the bubbles at the final time form a densely packed cluster
and $99.9958\%$ of cells contain four interfaces or less,
we choose to use~$L=4$ layers for all other simulations.

\subsection{Drop impact on liquid-liquid interface}

This test validates the model against experimental data on the gravity-driven
impact of a liquid drop onto a liquid-liquid interface and shows that our
method gives the same results as the multi-marked volume-of-fluid
method~\cite{coyajee2009}.
The problem is solved in a rectangular domain
of size $5\times10\times5\;\text{cm}$
on a mesh containing $160\times320\times160$ cells
bounded by no-slip walls in the vertical direction and periodic
conditions in the other directions.
Parameters of the problem are taken from numerical study~\cite{coyajee2009}
based on experiment~\cite{mohamed2003} (Combination~1):
density $\rho_1=949$ and
$\rho_2=1128\;\text{kg}/\text{m}^3$,
viscosity $\mu_1=0.019$ and $\mu_2=0.0063\;\text{Pa}\cdot\text{s}$,
gravitational acceleration $g=9.8\;\text{m}/\text{s}^2$
and surface tension coefficient $\sigma=0.029\;\text{N}/\text{m}$.
They correspond to a water+glycerin drop falling in silicon oil.
A spherical drop of a radius~$5.1~\text{mm}$ is initially placed
at a distance of $67$~\text{mm} between its center and the interface.

\begin{figure}[H]
  \centering
  \includegraphics{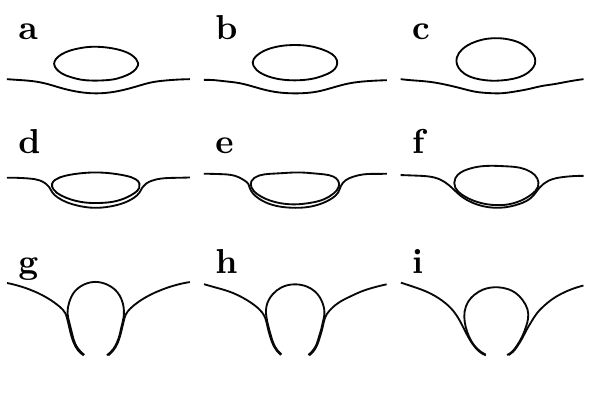}
  \caption{
    Drop impact on liquid-liquid interface. Central slices
    at $t=0.63,\;0.67,\;\text{and}\;0.77\;\text{s}$
    produced by the present method (\figbf{a,d,g})
    compared to images from numerical study~\cite{coyajee2009} (\figbf{b,e,h})
    and experimental data~\cite{mohamed2003} (\figbf{c,f,i}).
  }
  \label{f_dropimpact_slice}
\end{figure}

\suppreffigure{f_dropimpact_slice} compares the slices of the interface obtained
using the present method with numerical data~\cite{coyajee2009} and
experimental data~\cite{mohamed2003}.
Then the liquid film between the liquids drains but no coalescence occurs,
The falling drop approaches the surface separating the liquids,
creates a bulge and eventually rests on the surface.
Our algorithm produces the same results as method~\cite{coyajee2009} and both
agree with the experimental data.

\begin{figure}[H]
  \centering
  \includegraphics{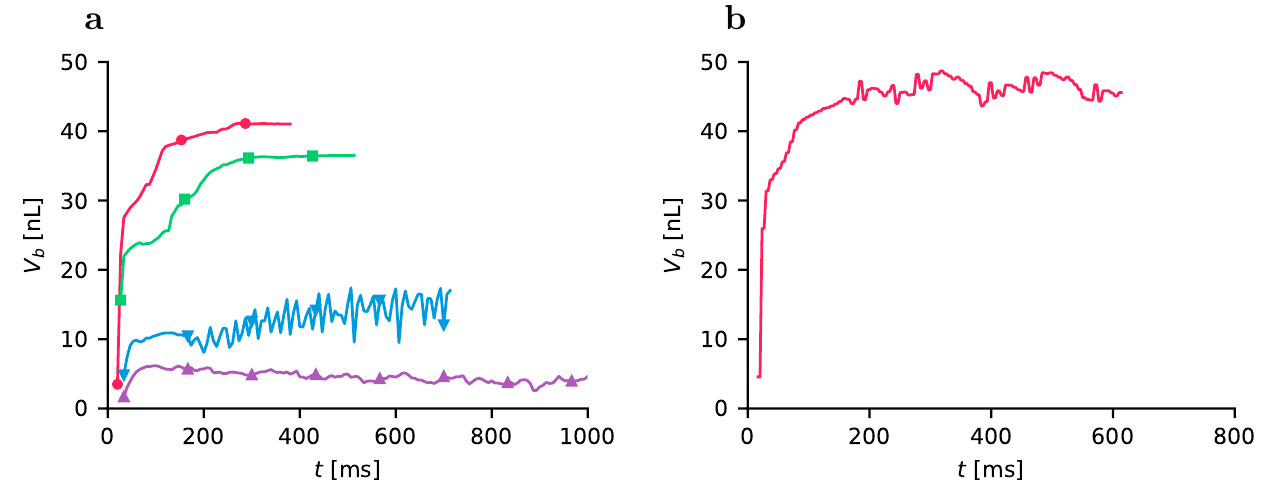}
  \caption{
    Microfluidic crystals.
    Evolution of the volume of generated bubbles.
    (\figbf{a}) Stationary crystalline structures
    in channel of length~$L=6.8\;\mathrm{mm}$.
    Hex-one ($P_g=216\;\mathrm{Pa}$)~\key{111},
    hex-two ($P_g=194\;\mathrm{Pa}$)~\key{122},
    hex-three ($P_g=149\;\mathrm{Pa}$)~\key{133},
    and hex-four ($P_g=140\;\mathrm{Pa}$)~\key{144}.
    (\figbf{b}) Spontaneous transitions between hex-one and hex-two
    at $P_g=225\;\mathrm{Pa}$ and the channel length~$L=5.3\;\mathrm{mm}$.
  }
  \label{f_crystal_volume}
\end{figure}

\begin{table}[H]
  \centering
  \begin{tabular}{lccc}
  density of liquid & $\rho_l$ & 1000 & $\mathrm{kg} \,/\, \mathrm{m}^3$ \\
  density of gas & $\rho_g$ & 100 & $\mathrm{kg} \,/\, \mathrm{m}^3$ \\
  viscosity of liquid & $\mu_l$ & 0.25 & $\mathrm{mPa} \cdot \mathrm{s}$ \\
  viscosity of gas & $\mu_g$ & 0.025 & $\mathrm{mPa} \cdot \mathrm{s}$ \\
  surface tension & $\sigma$ & 3.6 & $\mathrm{mN} \,/\, \mathrm{m}$ \\
  liquid flow rate & $Q_l$ & 0.3 & $\mathrm{mL} \,/\, \mathrm{h}$ \\
  bubble breakup period & $T_b$ & $0.5\,W_0^2H/Q_l$ & \\
  device height & $H$ & 100 & $\mu \mathrm{m}$ \\
  orifice width & $W_0$ & 110 & $\mu \mathrm{m}$ \\
  collection channel width & $W$ & 1000 & $\mu \mathrm{m}$ \\
  collection channel length & $L$ & 6800 & $\mu \mathrm{m}$ \\
  computational cell size & $h$ & 10.1 & $\mu \mathrm{m}$ \\
  &&&\\
  Case of bubbling oscillator: &&&\\
  collection channel length & $L$ & 5300 & $\mu \mathrm{m}$ \\
  bubble breakup period & $T_b$ & $0.46\,W_0^2H/Q_l$ & \\
    inlet pressure & $P_g$ & 225 & $\mathrm{Pa}$ \\
  \end{tabular}
  \caption{Microfluidic crystals. Simulation parameters.}
  \label{t_crystal}
\end{table}

\begin{table}[H]
  \centering
    \begin{tabular}{lccc}
  density of liquid & $\rho_l$ & 1021 & $\mathrm{kg} \,/\, \mathrm{m}^3$ \\
  density of gas & $\rho_g$ & 41 & $\mathrm{kg} \,/\, \mathrm{m}^3$ \\
  viscosity of liquid & $\mu_l$ & 1.25 & $\mathrm{mPa} \cdot \mathrm{s}$ \\
  viscosity of gas & $\mu_g$ & 0.05 & $\mathrm{mPa} \cdot \mathrm{s}$ \\
  surface tension & $\sigma$ & 35.2 & $\mathrm{mN} \,/\, \mathrm{m}$ \\
  bubble generation period & $T$ & 111 & $\mu \mathrm{s}$ \\
  volume of generated bubble & $V_b$ & 0.68 & $\mathrm{nL}$ \\
  gas volume fraction & $\Phi$ & 0.645 & $-$ \\
  inlet mean velocity & $v_\mathrm{{mean}}$ & 2.09 & $\mathrm{m} \,/\, \mathrm{s}$ \\
  inlet flow rate & $Q$ & 12 & $\mathrm{mL} \,/\, \mathrm{h}$ \\
  channel height & $H$ & 60 & $\mu \mathrm{m}$ \\
  inlet channel width & $W_0$ & 75 & $\mu \mathrm{m}$ \\
  inlet channel length & $L_0$ & 816 & $\mu \mathrm{m}$ \\
  collection channel width & $W$ & 1600 & $\mu \mathrm{m}$ \\
  collection channel length & $L$ & 1600 & $\mu \mathrm{m}$ \\
  computational cell size & $h$ & 2.51 & $\mu \mathrm{m}$ \\
  \end{tabular}
  \caption{Bidisperse foam generation. Simulation parameters.}
  \label{t_microfoam}
\end{table}

\begin{table}[H]
  \centering
  \begin{tabular}{lcc}
  density of liquid & 1000 & $\mathrm{kg} \,/\, \mathrm{m}^3$ \\
  density of gas & 30 & $\mathrm{kg} \,/\, \mathrm{m}^3$ \\
  viscosity of liquid & $2\cdot10^{-3}$ & $\mathrm{Pa} \cdot \mathrm{s}$ \\
  viscosity of gas & $2\cdot10^{-5}$ & $\mathrm{Pa} \cdot \mathrm{s}$ \\
  surface tension & 18 & $\mathrm{mN} \,/\, \mathrm{m}$ \\
  bubble diameter & 2 & $\mathrm{mm}$ \\
  domain length & 30 & $\mathrm{mm}$ \\
  domain width & 30 & $\mathrm{mm}$ \\
  domain height & 20 & $\mathrm{mm}$ \\
  computational cell size & 0.16 & $\mathrm{mm}$ \\
  \end{tabular}
  \caption{Clustering of bubbles. Simulation parameters.}
  \label{t_column}
\end{table}

\begin{table}[H]
  \centering
  \begin{tabular}{lcc}
  density of liquid & 1000 & $\mathrm{kg} \,/\, \mathrm{m}^3$ \\
  density of gas & 10 & $\mathrm{kg} \,/\, \mathrm{m}^3$ \\
  viscosity of liquid & $10^{-3}$ & $\mathrm{Pa} \cdot \mathrm{s}$ \\
  viscosity of gas & $10^{-5}$ & $\mathrm{Pa} \cdot \mathrm{s}$ \\
  surface tension & 72 & $\mathrm{mN} \,/\, \mathrm{m}$ \\
  waterfall mean velocity & 1.5 & $\mathrm{m}/\mathrm{s}$ \\
  waterfall thickness & 5 & $\mathrm{mm}$ \\
  domain length & 200 & $\mathrm{mm}$ \\
  domain width & 100 & $\mathrm{mm}$ \\
  domain height & 100 & $\mathrm{mm}$ \\
  \end{tabular}
  \caption{Foaming waterfall. Simulation parameters.}
  \label{t_waterfall}
\end{table}

\begin{figure}[H]
  \centering
  \includegraphics{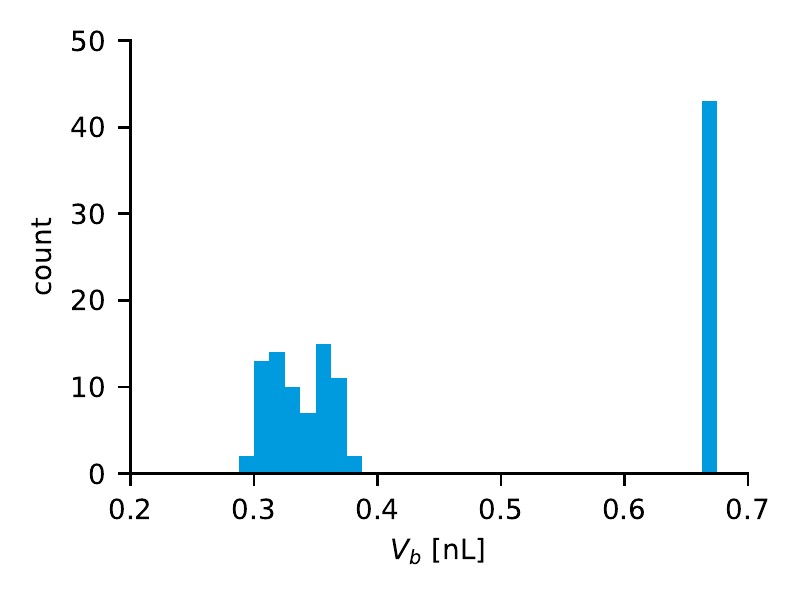}
  \caption{
    Bidisperse foam generation.
    Histogram of the bubble volume after 80 cycles.
  }
  \label{f_microfoam_hist}
\end{figure}

\begin{figure}[H]
  \centering
  \includegraphics{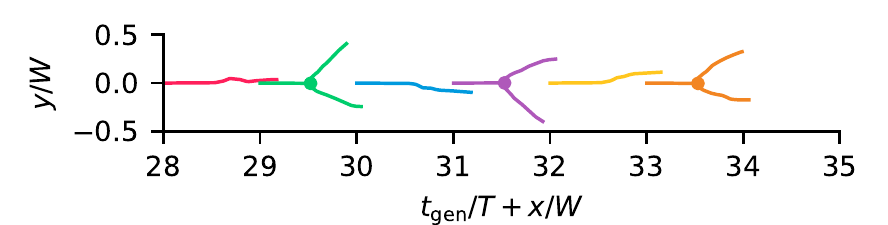}
  \caption{
    Bidisperse foam generation.
    Trajectories of the centroids of bubbles generated
    at~$t_\mathrm{gen}/T\in[28, 34)$
    shifted along the horizontal axis by the generation time.
    Coordinates of the centroid~$x$ and $y$ are relative to the generation point.
  }
  \label{f_microfoam_events}
\end{figure}

\begin{figure}[H]
  \centering
  \includegraphics{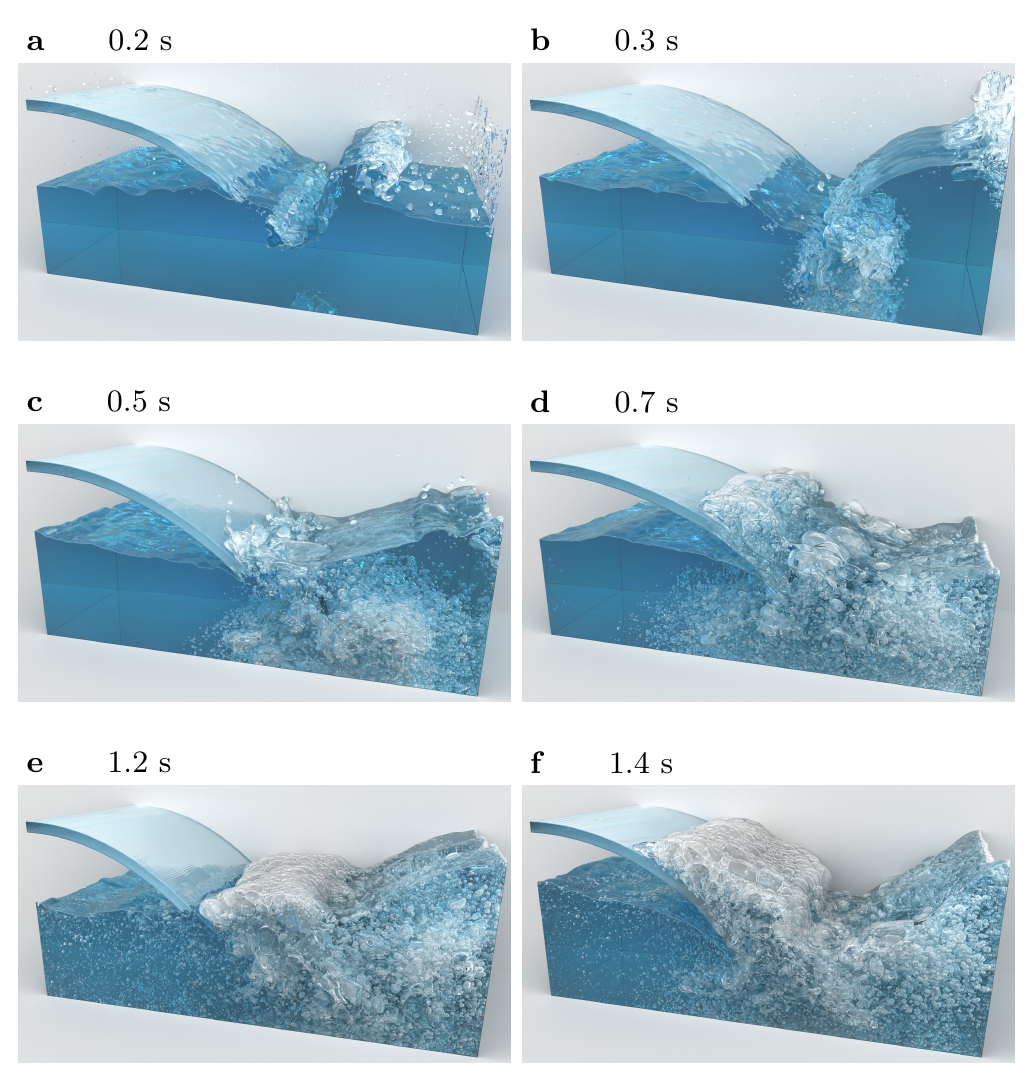}

  \caption{
    Foaming waterfall.
    Snapshots of the interface at
    $t=0.2$~(\figbf{a}),
    $0.3$~(\figbf{b}),
    $0.5$~(\figbf{c}),
    $0.7$~(\figbf{d}),
    $1.2$~(\figbf{e}), and
    $1.4~\mathrm{s}$~(\figbf{f}).
  }
  \label{f_waterfall_series}
\end{figure}

\bibliography{main}